\documentclass[a4paper,12pt]{article}
\pdfoutput=1

\usepackage[a4paper,left=80pt,right=80pt,top=78pt,bottom=80pt]{geometry}
\usepackage[latin1]{inputenc}
\usepackage{dsfont}
\usepackage{amsfonts,amsmath,amssymb}
\usepackage{amsthm}
\usepackage{graphicx}
\usepackage[small,bf]{caption}
\usepackage{subcaption}
\usepackage{booktabs}
\usepackage[numbers, sort&compress]{natbib}
\allowdisplaybreaks[1]
\usepackage{color}
\usepackage{ulem}
\usepackage{tikz}
\usepackage{multirow}
\usetikzlibrary{arrows}
\usepackage{xspace}
\usepackage{subfig}
\usepackage{enumitem}

\def\bal#1\eal{\begin{align}#1\end{align}}
\newcommand{\be}{\begin{equation}}
\newcommand{\ee}{\end{equation}}
\newcommand{\bea}{\begin{eqnarray}}
\newcommand{\eea}{\end{eqnarray}}
\newcommand{\besub}{\begin{subequations}}
\newcommand{\eesub}{\end{subequations}}
\newcommand{\ba}{\begin{array}}
\newcommand{\ea}{\end{array}}
\newcommand{\bi}{\begin{itemize}}
\newcommand{\ei}{\end{itemize}}
\newcommand{\nn}{\nonumber}

\newcommand{\Lcal}{{\cal L}}

\newcommand{\hc}{{\textrm{h.c.}}}

\begin{document}

\begin{titlepage}

\flushright{ 
TUM-HEP-1072-16
 }
 
\vspace*{1.0cm}

\begin{center}
{\LARGE 
{\bf
Evading Direct Dark Matter  
 Detection \\ in Higgs Portal Models
}}
\\
[1.0cm]

{
{\bf
Giorgio Arcadi$^{1}$, Christian Gross$^{2}$, Oleg Lebedev$^{2}$,\\  
 Stefan Pokorski$^{3}$, Takashi Toma$^{4}$
}}
\end{center}
%\addtocounter{footnote}{-3}
\vspace*{0.3cm}

\centering{

$^{1}$ 
\it{
Max Planck Institut f\"{u}r Kernphysik, Saupfercheckweg 1,\\
D-69117 Heidelberg, Germany
}

$^{2}$ 
\it{Department of Physics and Helsinki Institute of Physics, \\
Gustaf H\"allstr\"omin katu 2, FI-00014 Helsinki, Finland
}

$^{3}$ 
\it{Institute of Theoretical Physics, University of Warsaw, \\
Pasteura 5, PL-02-093 Warsaw, Poland
}

$^{4}$ 
\it{
Physik-Department T30d, Technische Universit\"at M\"unchen,\\
 James-Franck-Stra\ss{}e, D-85748 Garching, Germany
}
}

\vspace*{1.0cm}

\begin{abstract}
Many models of Higgs portal Dark Matter (DM) find themselves under pressure from increasingly tight direct detection constraints. In the framework of gauge field DM, we study how such bounds can be relaxed while retaining the thermal WIMP paradigm. When the hidden sector gauge symmetry is broken via the Higgs mechanism, the hidden sector generally contains unstable states which are lighter than dark matter. These states provide DM with an efficient annihilation channel. As a result, the DM relic abundance and the direct detection limits are controlled by different parameters, and the two can easily be reconciled. This simple setup realizes the idea of ``secluded'' dark matter naturally.
\end{abstract}

%\today

\end{titlepage}
\newpage

%\tableofcontents

%=========================================================================
%=========================================================================
\section{Introduction}
%=========================================================================
%=========================================================================

The Higgs sector of the Standard Model (SM) enjoys a special feature that
it can couple to the hidden sector at the renormalizable level.
In particular, a ``Higgs portal'' interaction term
\cite{Silveira:1985rk,Schabinger:2005ei,Patt:2006fw}
\begin{equation}
V_{\rm portal}=\lambda_{h\phi} \vert H\vert^2 \vert \phi \vert^2 \;,
\label{eq1}
\end{equation}
where $\phi$ is a hidden sector scalar, is allowed by all symmetries and has dimension 4. Thus, interactions of this type are expected on general grounds. 

An interesting application of this observation is that 
the Higgs field can couple to dark matter (DM), which is thought to reside in the hidden sector. If the hidden sector is endowed with gauge symmetry, a natural DM candidate would be the corresponding vector field \cite{Hambye:2008bq,Lebedev:2011iq,Gross:2015cwa}. Indeed, the U(1) and SU(N) spontaneous symmetry breaking 
with a minimal number of scalar fields implies stability of some of the massive gauge fields. This is due to a residual symmetry which acts
on the hidden sector states only. In the most general case, multi--scalar systems break CP such that the stabilizing symmetry is
$Z_2$ (or a generalization thereof),
\begin{equation}
A_\mu^a \rightarrow (-1)^{n_a} A_{\mu}^a \;,
\end{equation}
where $a$ is a group index and $n_a$ is an integer. In the non--Abelian
case, this $Z_2$ is part of a larger unbroken group, e.g. SO(3) or U(1) \cite{Hambye:2008bq,Arcadi:2016kmk}. It should be noted that for U(1) and SU(2) gauge groups, CP is unbroken since only a single field is required to break the symmetry
and the above $Z_2$ can be viewed as charge conjugation. 

The hidden sector ``Higgs'' field(s) $\phi$ mixes with the SM Higgs
due to the portal coupling Eq.~(\ref {eq1}). Therefore, the 125 GeV scalar 
couples to DM, although such a coupling is suppressed by the mixing angle. As a result, the hidden sector DM can annihilate into the SM states and scatter off nuclei as a conventional WIMP would. 
The current direct DM detection constraints from LUX and PANDA experiments \cite{Akerib:2015rjg,Tan:2016zwf} are so tight that the WIMP paradigm within the Higgs portal
framework finds itself under pressure. The core of the problem is that the couplings controlling DM annihilation and its scattering off nuclei are related, while the direct detection bound requires the latter to be small.

In this work, we emphasize that the Higgs portal models with gauged
hidden sectors possess unstable states which can be lighter than
dark matter.
 This provides DM with an efficient annihilation channel
which breaks the correlation between the annihilation cross section and
the direct detection rate. Thus, all of the constraints can easily be satisfied in this kinematic regime. 
This type of DM is known as a ``secluded WIMP'' \cite{Pospelov:2007mp,Pospelov:2008jd}, while in this work
we show that it is realized quite naturally in the Higgs portal models
with gauged hidden sectors. In the context of an SU(2) model,
this phenomenon was noted in \cite{Hambye:2008bq}. 
The same idea applies of course 
to non--gauge hidden sectors as long as there are unstable fields lighter than DM
(see e.g. \cite{LopezHonorez:2012kv}).

%=========================================================================
\section{Hidden U(1) sector}
%=========================================================================

The simplest example of a vector DM model with a natural $Z_2$ symmetry \cite{Hambye:2008bq,Lebedev:2011iq} (see also \cite{Farzan:2012hh,Baek:2012se,Duch:2015jta}) is a hidden Abelian gauge sector.
Within effective field theory, the model (with a heavy ``hidden Higgs'') was analyzed in
 \cite{Kanemura:2010sh,Djouadi:2011aa}.

 The Lagrangian is given by
 \begin{equation}
 {\cal L_{\rm hidden}}= -{1\over 4} F_{\mu\nu} F^{ \mu\nu} + (D_\mu \phi)^\dagger D^\mu \phi -V(\phi) \;,
 \end{equation}
 where $\phi$ is a charged scalar, $F_{\mu\nu}$ is the U(1) field strength of the gauge field $A_\mu$ and $V(\phi)$ is the potential. 
We take the charge of $\phi $ to be +1/2 for easier $ $comparison to the non--Abelian case.
In unitary gauge $\phi$ can be written as $\phi=(\tilde v + \rho)/\sqrt{2}$ where $\tilde v$ is the VEV and $\rho$ a real scalar field.
The imaginary part of $\phi$ is eaten by $A_\mu$ which obtains the mass $m_A= \tilde g \tilde v /2$, where $\tilde g$ is the gauge coupling.
The gauge--scalar interactions are given by
 \begin{eqnarray}
&& \Delta {\cal L}_{\rm s-g}= {\tilde g^2\over 4} \tilde v \rho \; A_\mu A^{ \mu} +
 {\tilde g^2\over 8} \rho^2 \; A_\mu A^{\mu} \;. 
\label{rho-A} 
 \end{eqnarray}
The $Z_2$ symmetry 
\begin{equation}
A_\mu \rightarrow - A_\mu ~,
\end{equation} 
which is the usual charge $ $conjugation $ $symmetry, makes the massive gauge field stable so that the latter is a viable dark matter candidate.

The visible and hidden sectors interact via the Higgs portal coupling
\begin{equation}
 {\cal L_{\rm portal}}= - \lambda_{h \phi} \vert H \vert^2 \vert \phi \vert^2 \;.
 \end{equation}
This coupling leads to the mixing of $\rho$ with the Higgs, which in unitary gauge can be written as $H^T= (0, v+h)/\sqrt{2}$.
The fields $\rho$ and $h$ can be written in terms of mass eigenstates $h_{1,2}$ as
\begin{eqnarray}
&& \rho= - h_1 \; \sin\theta + h_2 \; \cos\theta \;, \nonumber\\
&& h = h_1 \; \cos\theta + h_2 \; \sin\theta \;,
 \end{eqnarray}
where $\theta$ is the Higgs mixing angle and we identify $h_1$ with the 125 GeV Higgs.

Here we assume that 
the tree level kinetic mixing between the hypercharge gauge boson and $A_\mu$ is zero. This happens if the corresponding generators are orthogonal in the UV completion. For instance, the observable sector
can originate from one $E_8$ factor of the $E_8 \times E_8$ string theory, while the hidden sector comes from the other~\cite{Gross:1984dd}. The kinetic mixing is not generated radiatively as long as the interaction between the two sectors is due to the Higgs portal term.

Let us now discuss the main phenomenological features of this scenario. 
All the relevant scattering processes, including DM annihilation and DM scattering on nucleons, proceed through $h_1$ and $h_2$ exchange.
The dark matter--nucleon interaction cross section is given by
(see e.g. \cite{Gross:2015cwa})
\begin{equation}
\sigma^{\rm SI}_{A-N}= {g^2 \tilde g^2 \over 16 \pi} \; {m_N^2 \mu_{\rm A N}^2 f_N^2 \over m_W^2} \;
{ (m_{h_2}^2 - m_{h_1}^2 )^2 \sin^2 \theta \; \cos^2 \theta \over m_{h_1}^4 m_{h_2}^4} \;,
\end{equation}
where $m_N$ is the nucleon mass, $\mu_{\rm A N}=m_A m_N/(m_A + m_N)$ and $f_N \simeq 0.3$ parametrizes the Higgs--nucleon coupling. One should keep in mind that there is an uncertainty in $f_N$ and here
we use the default micrOMEGAs \cite{Belanger:2014vza} value.

\begin{figure}[h]
\begin{center}
\includegraphics[width=6.8 cm]{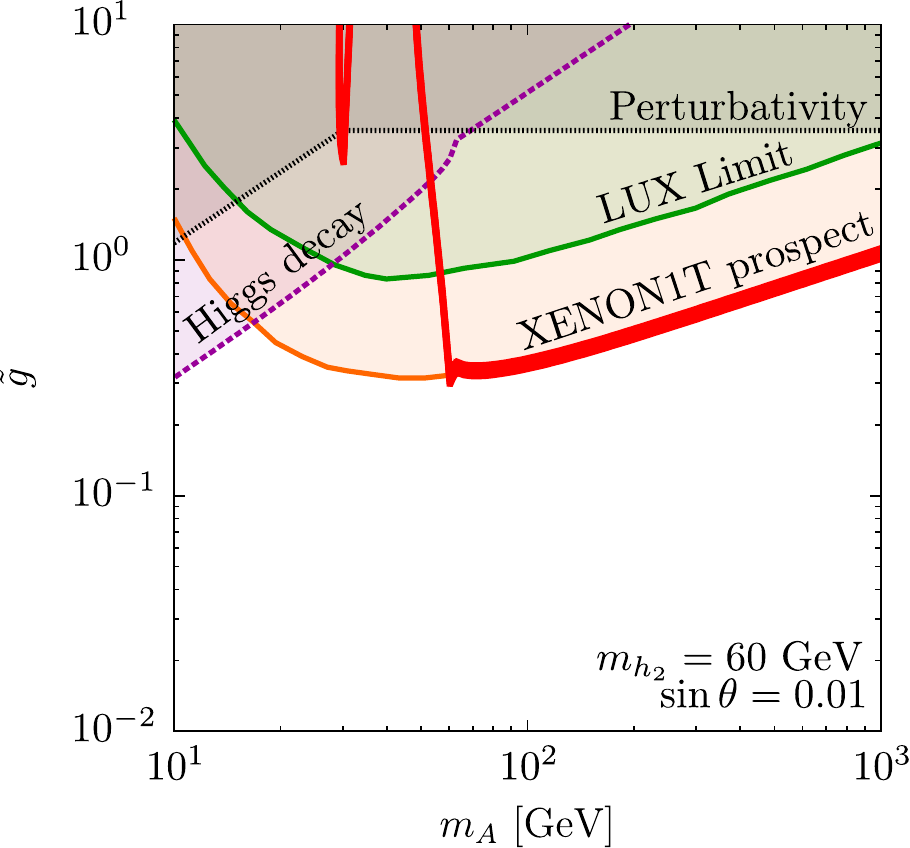}
\includegraphics[width=6.8 cm]{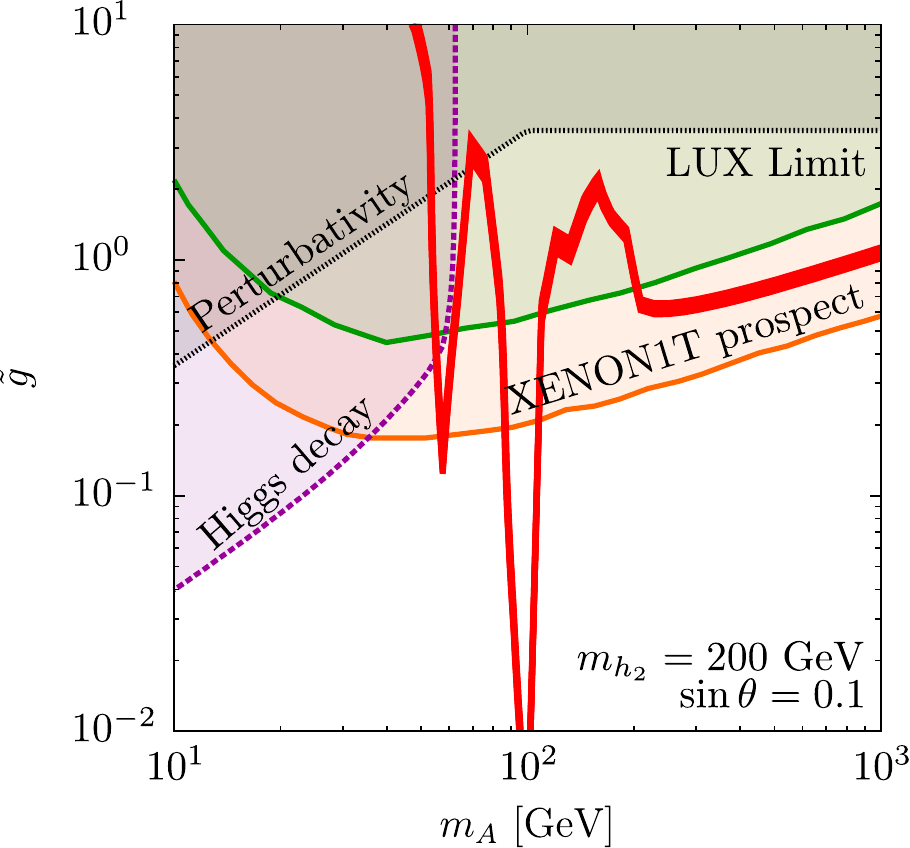}
\hspace{0.3cm}
\includegraphics[width=6.8 cm]{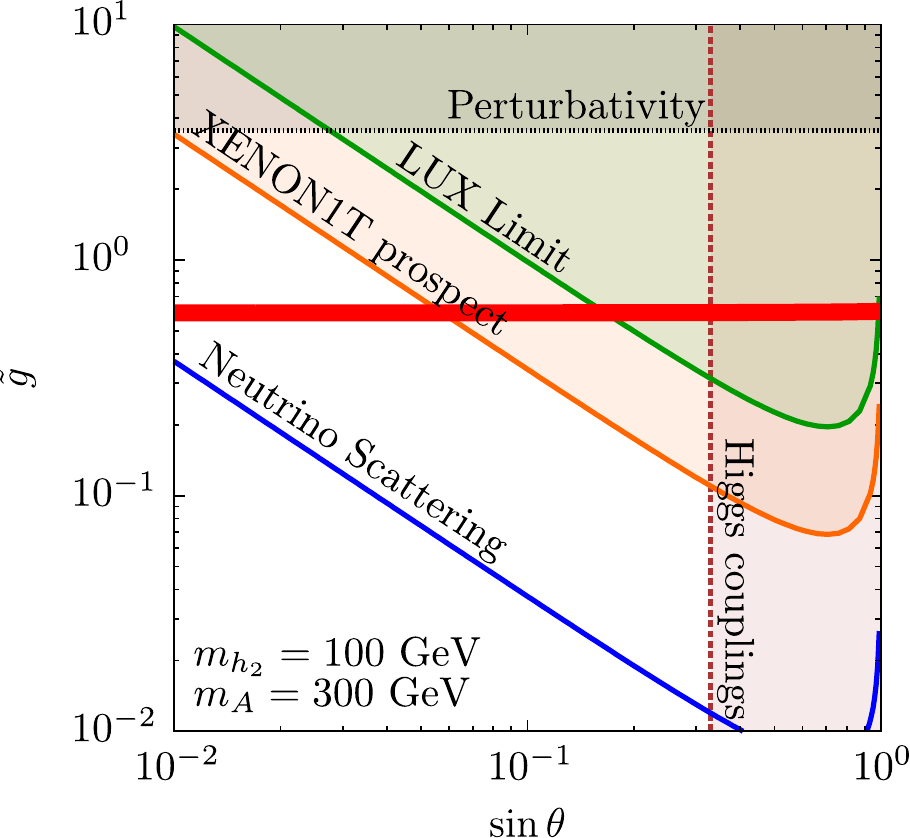}
\includegraphics[width=6.8 cm]{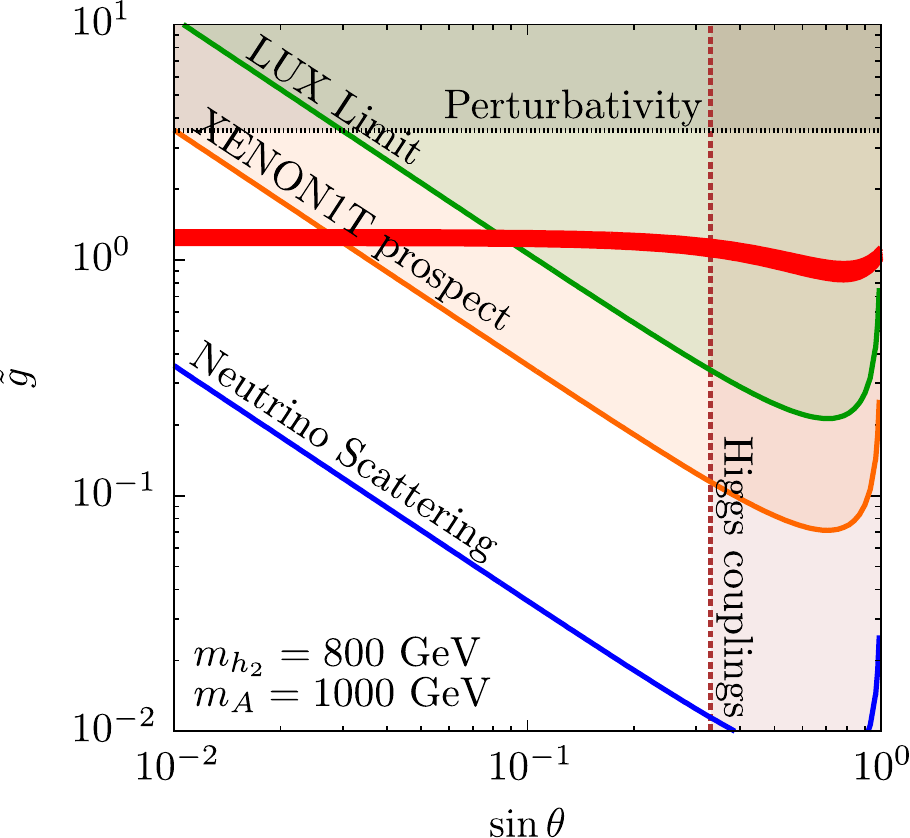}
\end{center}
\caption{ Dark matter constraints
 in the plane $(m_A,\tilde{g})$ (upper panels) and $(\sin\theta,\tilde{g})$ (lower panels) for U(1) DM. 
The red band indicates the correct relic DM density. The other curves mark the following constraints:
grey -- perturbativity, purple -- invisible Higgs decay, dark red -- Higgs couplings, green -- LUX 2016 direct DM detection, orange -- XENON1T direct DM detection prospects. The blue line represents the direct detection event rate corresponding to the relic neutrino scattering off nuclei.
 }
\label{fig:dm1}
\end{figure}

 With regard to dark matter annihilation,
 we focus on the kinematic regime $m_A > m_{h_2}$. For a small $\sin\theta$, 
 the main annihilation channel in our study is $AA\rightarrow h_2 h_2$ since the relevant vertices are not $\theta$--suppressed.
The contributions to this process include the $t$-- channel $A$--exchange, the $s$--channel $h_{1,2}$
exchange as well as the contact $AAh_2h_2$ term. The full
 cross section expression is quite bulky and not particularly illuminating, thus let us only quote the limit 
 $\sin\theta \ll 1$ and $m_{h_1} \ll m_A,m_{h_2}$,
\bal
 \langle\sigma{v}\rangle=
\frac{\tilde{g}^4}{576\pi m_A^2}\sqrt{1-\frac{m_{h_2}^2}{m_A^2}}
\frac{11m_{h_2}^8-80m_{h_2}^6m_A^2+240m_{h_2}^4m_A^4-320m_{h_2}^2m_A^6+176m_A^8}
{\left(4m_A^2-m_{h_2}^2\right)^2\left(2m_A^2-m_{h_2}^2\right)^2}.
\eal
 In our numerical studies, however, we use the exact result.

In Fig.~\ref{fig:dm1}, we display the results of our numerical studies
using the package Micromegas \cite{Belanger:2014vza}.
Apart from the direct detection and relic density constraints, we show the perturbativity bounds for  the gauge and scalar couplings. 
In the  upper panels, the Higgs decay constraint is also displayed. 
It comes from the requirement that the LHC Higgs signal strength $\mu$  be 
close to the corresponding SM prediction. 
 When the Higgs production is approximately SM--like, as is the case for small $\sin\theta$, the experimental result $\mu= 1.09^{+0.11}_{-0.10}$~\cite{Khachatryan:2016vau} translates into $\textrm{BR}(h \to \textrm{invisible}) \leq 0.11$ at 95\% CL. This constrains the $h_1$ coupling to DM
 and to $h_2$ when the corresponding decay channels are open. 
For the parameter choices of the   lower panels, this bound is not relevant. Instead, we display there an upper bound on    $\sin \theta$ from the Higgs coupling
measurements. The mixing between the SM Higgs and the hidden Higgs suppresses all the
couplings of the 125 GeV scalar universally, and is thus subject to the strong LHC bounds.

The thermal WIMP paradigm is consistent with the LUX constraint for
$m_A > m_{h_2}$ (plus the resonance regions $m_A \simeq m_{h_{1,2}}/2$) and a sufficiently small $\theta$. If $h_2$ is relatively heavy,
$m_{h_2}\sim {\cal O}(100)$ GeV, the required mixing angle is about 0.1 or less. For a light $h_2$, the direct detection rate gets quite high and a smaller $\theta \sim {\cal O}(10^{-2})$ is necessary. Note also that
in the region $m_{h_1} \sim m_{h_2}$, there are substantial cancellations in $\sigma_{A-N}$.

The lower panels show that the annihilation cross section at 
$m_A > m_{h_2}, m_{h_1}$ is largely independent of $\sin \theta$.
This is because, neglecting the $h_1-h_2$ mass difference, the gauge boson interactions with $h_{1,2}$ are equivalent to those with $\rho$
(Eq.~\ref{rho-A}) which are independent of $\theta$. The $s$--channel
diagrams with $h_{1,2} $--exchange include $\theta$--dependent Higgs vertices. These however are subleading such that the full 
cross section depends on $\theta$ very weakly.
We see from Fig.~\ref{fig:dm1} that the upper bound on $\sin \theta$ from
LUX is about 0.1 for the chosen $m_{h_2}$. Our dark matter candidate
evades XENON1T detection for somewhat smaller $\sin\theta $ values,
between $10^{-2}$ and $10^{-1}$.

These plots illustrate very well our main point: in the regime 
$m_A > m_{h_2}$, the direct detection rate is almost uncorrelated with the annihilation cross section.

The SU(2) hidden sector case \cite{Hambye:2008bq,Khoze:2014xha,Karam:2015jta} is very similar to the Abelian case considered here since both symmetries can be broken by a single field.
For larger groups, the situation is more involved and in the following section we study an SU(3) example.

\section{Hidden SU(3) sector}

For SU(3) and larger groups \cite{Gross:2015cwa},
the dark matter composition depends on whether CP is broken in the scalar sector. In general, the scalar potential with multiple fields 
allows for complex couplings which violate CP. In what follows,
we will adhere to this general situation, in which case DM is composed
of gauge fields.\footnote{If CP is preserved, DM may have a pseudoscalar component \cite{Arcadi:2016kmk}.} The corresponding stabilizing symmetry is a global
subgroup of the gauge group.

Let us now consider in detail a hidden sector endowed with SU(3)
gauge symmetry following Ref.~\cite{Gross:2015cwa}.
The symmetry is broken spontaneously (to nothing) by VEVs of two triplets
$\phi_1$ and $\phi_2$. This is the minimal setup that allows one to 
make all the SU(3) gauge fields massive. A variation of this model
has been considered in \cite{Karam:2016rsz}.

The Lagrangian of the model is
\besub
\bal
-\Lcal_{\rm portal} & = \lambda_{H11} \, |H|^2 |\phi_1|^2 + \lambda_{H22} \, |H|^2 | \phi_2|^2 - ( \lambda_{H12} \, |H|^2 \phi_1^\dagger \phi_2 +\hc)\;,
 \\
\Lcal_{\rm hidden} &= - \frac12 \textrm{tr} \{G_{\mu \nu} G^{\mu \nu}\} + |D_\mu \phi_1|^2 + |D_\mu \phi_2|^2 -V_{\rm hidden} \,,
\eal
\eesub
where $G_{\mu \nu}=\partial_\mu A_\nu - \partial_\nu A_\mu + i \tilde g [A_\mu,A_\nu]$ is the field strength of the SU(3) gauge fields $A_\mu^a$, $\tilde g$ is the gauge coupling, $D_\mu \phi_i = \partial_\mu \phi_i + i \tilde g A_{\mu} \phi_i$ is the covariant derivative of $\phi_i$ and $H$ is the Higgs doublet, which in the unitary gauge can be written as $H^T= (0,v+h)/\sqrt{2}$.
The most general $ $renormalisable $ $hidden $ $sector potential can be written as
\bal
V_{\rm hidden}(\phi_1,\phi_2) &=
m_{11}^2 |\phi_1|^2
+ m_{22}^2 |\phi_2|^2
- ( m_{12}^2 \phi_1^\dagger \phi_2 + \hc )
\nn \\ 
& 
+ \frac{\lambda_1}{2} |\phi_1|^4
+ \frac{\lambda_2}{2} |\phi_2|^4
+ \lambda_3 |\phi_1|^2 |\phi_2|^2
+ \lambda_4 | \phi_1^\dagger\phi_2 |^2
\nn \\ 
& 
+ \left[
\frac{ \lambda_5}{2} ( \phi_1^\dagger\phi_2 )^2
+ \lambda_6 |\phi_1|^2
( \phi_1^\dagger\phi_2)
+ \lambda_7 |\phi_2|^2
( \phi_1^\dagger\phi_2 )
+ \hc \right] \,.
\label{V}
\eal
In the $ $unitary $ $gauge, the fields $\phi_1$ and $\phi_2$  which are responsible for spontaneous SU(3) breaking (to nothing), can be written as
\be \label{unitarygauge}
\phi_1={1\over \sqrt{2}} \,
\left( \begin{array}{c}
0\\0\\v_1+\varphi_1
\end{array} \right) \,,
\quad 
\phi_2= {1 \over \sqrt{2}}\,
\left( \begin{array}{c}
0\\v_2+\varphi_2\\v_3+\varphi_3 + iv_4+ i \varphi_4
\end{array} \right) ~,
\ee
where the $v_i$ are VEVs and $\varphi_{i}$ are real scalar fields. 
In general, CP is broken in the scalar sector and 
 all of the scalar fields mix. In what follows, we make a simplifying assumption that $v_3$ and $v_4$ are small and can be neglected. This
 makes the analysis more tractable without affecting the essence of
 the model. For definiteness, we also take $v_1 >v_2$.

Our setup enjoys a symmetry that makes the fields 
 $A^1_\mu$ and
$A^2_\mu$ stable. To see this, it is sufficient to realize that
the model is symmetric under $A^{1,2}_\mu \rightarrow
-A^{1,2}_\mu$. This reflection symmetry is part of an unbroken global U(1) 
which corresponds to the SU(3) gauge transformation $U A_\mu U^\dagger$ with 
\begin{equation}
U= e^{ i \xi/3} \;{\rm diag} (e^{- i \xi} , 1 , 1)\;. 
\label{U1}
\end{equation}
Under this symmetry, the gauge field components are rotated as
$A^{1(2)} \rightarrow A^{1(2)} \cos\xi \mp A^{2(1)} \sin\xi $
and $A^{4(5)} \rightarrow A^{4(5)} \cos\xi \mp A^{5(4)} \sin\xi $.
Thus only $A^{1,2,4,5}$ have non--trivial U(1) quantum numbers,
while $A^{3,6,7,8}$ are neutral and can decay into SM matter.
The scalar sector has an independent global U(1)$^\prime$ symmetry $\phi_{1,2} \rightarrow e^{i \beta}\;\phi_{1,2}$. 
Since $U$ acts effectively as an overall phase transformation on the scalar fields Eq.~(\ref{unitarygauge}), the vacuum preserves a combination of U(1)$^\prime$ and $U$. 
This symmetry has, in particular, the consequence that $m_{A_1}= m_{A_2}$ and $m_{A_4}= m_{A_5}$ (see \cite{Gross:2015cwa}).
It is intact as long as SU(3) is broken in the minimal fashion,
that is, via VEVs of only 2 triplets.

Details of the particle spectrum can be found in \cite{Gross:2015cwa,Arcadi:2016kmk}, while for our purposes it is sufficient to highlight 
the following features. 
In the limit $v_3,v_4 \ll v_1,v_2$ the vector sector is composed of 6 pure states which form 3 mass degenerate pairs with masses
\begin{equation}
\label{eq:vmasses_1}
m_{A^1}^2=m_{A^2}^2=\frac{\tilde{g}^2}{4}v_2^2,\quad
m_{A^4}^2=m_{A^5}^2=\frac{\tilde{g}^2}{4}v_1^2,\quad
m_{A^6}^2=m_{A^7}^2=\frac{\tilde{g}^2}{4}(v_1^2+v_2^2) \,,
\end{equation}
and two mixed eigenstates
\begin{align}
& A_\mu^{3\,'}=A_\mu^3 \cos\alpha + A_\mu^8 \sin\alpha~, \nonumber\\
& A_\mu^{8\,'}=A_\mu^8 \cos\alpha - A_\mu^3 \sin\alpha~,
\end{align}
where 
\be
\alpha=\left \{ 
\begin{array}{ll}
\frac12 \arctan \left( \frac{\sqrt{3}v_2^2}{2 v_1^2-v_2^2} \right)
&
\textrm{for} \quad v_2^2 \leq 2 v_1^2
\vspace{5pt}
\\
\frac12 \arctan \left( \frac{\sqrt{3}v_2^2}{2 v_1^2-v_2^2} \right) + \frac \pi 2
&
\textrm{for} \quad v_2^2 > 2 v_1^2
\end{array}
\right.
\ee
so that $\alpha \in (0^\circ, 60^\circ)$.
Their masses are
\begin{align}
\label{eq:vmasses_2}
& m_{A^{3\,'}}^2=\frac{\tilde{g}^2 v_2^2}{4}\left(1-\frac{\tan \alpha}{\sqrt{3}}\right),\,\,\,\,\,m_{A^{8\,'}}^2=\frac{\tilde{g}^2 v_1^2}{3}\frac{1}{1-\frac{\tan \alpha}{\sqrt{3}}} \,.
\end{align}
Since $\tan\alpha >0$, an important consequence of this formula is that 
\begin{equation}
m_{A^{1,2}} > m_{A^{3 \; \prime}} \;.
\label{AA3}
\end{equation}

The lightest fields with non--trivial U(1) quantum numbers are 
 $A^{1,2}_\mu$. They are stable and thus
 can play the role of dark matter. From now on, we will denote them by $A$ for brevity.
Other fields decay into either these fields plus SM states or entirely
into the SM final states.

Due to Eq.~(\ref{AA3}), 
the DM annihilation channel 
\begin{equation}
AA \rightarrow
A^{3\; \prime }A^{3\; \prime}
\end{equation}
 is $always$ open and does not suffer the $\sin\theta$ suppression.\footnote{In practice,  $A^{3\; \prime}$ is   slighter lighter than $A^{1,2}$ such that  this channel incurs some phase space suppression, yet remains efficient.}
 $A^{3\;\prime}$ is invariant under the transformation of Eq.~(\ref{U1}) and thus
 it decays into SM fields via off--shell scalars. Therefore the ``secluded'' DM scenario \cite{Pospelov:2007mp} is realized here naturally. The relevant interactions are 
\begin{eqnarray}
\mathcal{L}
\hspace{-0.2cm}&=&\hspace{-0.2cm}
\;\frac{\tilde{g}^2}{4}v_2\left(-s_\theta h_1+c_\theta h_2\right)
\left[
\sum_{a=1,2}A_{\mu}^{a}A^{a\mu}
+\left(\cos\alpha-\frac{\sin\alpha}{\sqrt{3}}\right)^2A_{\mu}^{3\prime}A^{3\prime\mu}
\right]\nonumber\\
\hspace{-0.2cm}&&\hspace{-0.2cm}
+\; \tilde{g}\cos\alpha 
\sum_{a,b,c=1,2,3^\prime} \epsilon_{abc}\; \partial_{\mu}A_{\nu}^{a}\; A^{b\mu}A^{c\nu}
\nonumber\\
\hspace{-0.2cm}&&\hspace{-0.2cm}
-\; \frac{\tilde{g}^2}{2}\cos^2\alpha \sum_{a=1,2} \left(
A_{\mu}^{a}A^{a\mu}\left(A_{\nu}^{3\prime}A^{3\prime\nu}\right)
-\left(A_{\mu}^{a}A^{3\prime\mu}\right)^2
\right),
\end{eqnarray}
where the $\epsilon_{abc}$ tensor is antisymmetric in indices $1,2,3^\prime$. In this expression, we have neglected contributions of heavier scalar and vector states (see the spectrum in \cite{Arcadi:2016kmk}).

 The analysis of the scalar sector is facilitated assuming small CP breaking. In that case, one can repeat the analysis
of \cite{Arcadi:2016kmk} while keeping in mind that all the scalars mix and therefore are unstable. Following \cite{Arcadi:2016kmk},
the lightest spin-0 state can be a ``mostly pseudoscalar'' $\chi$ closely related to $\varphi_4$. 
Depending on the parameter region, efficient DM annihilation
channels 
\begin{equation}
AA \rightarrow \chi\chi ~,~ h_2 h_2 
\end{equation}
can be available. These are unsuppressed by $\sin\theta$ and provide
a further mechanism to ``seclude'' \cite{Pospelov:2007mp} our dark matter. The corresponding analytical expressions are bulky and we omit
them in this work.
 
 \begin{figure}[h]
\begin{center}
\includegraphics[width=6.8 cm]{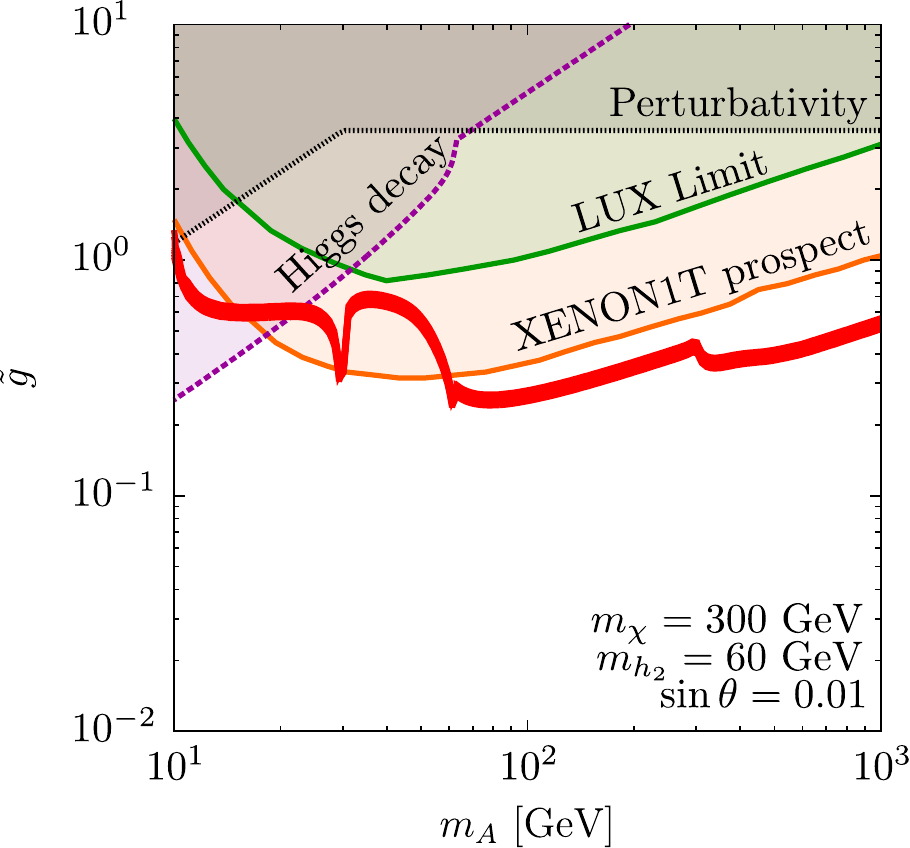}
\includegraphics[width=6.8 cm]{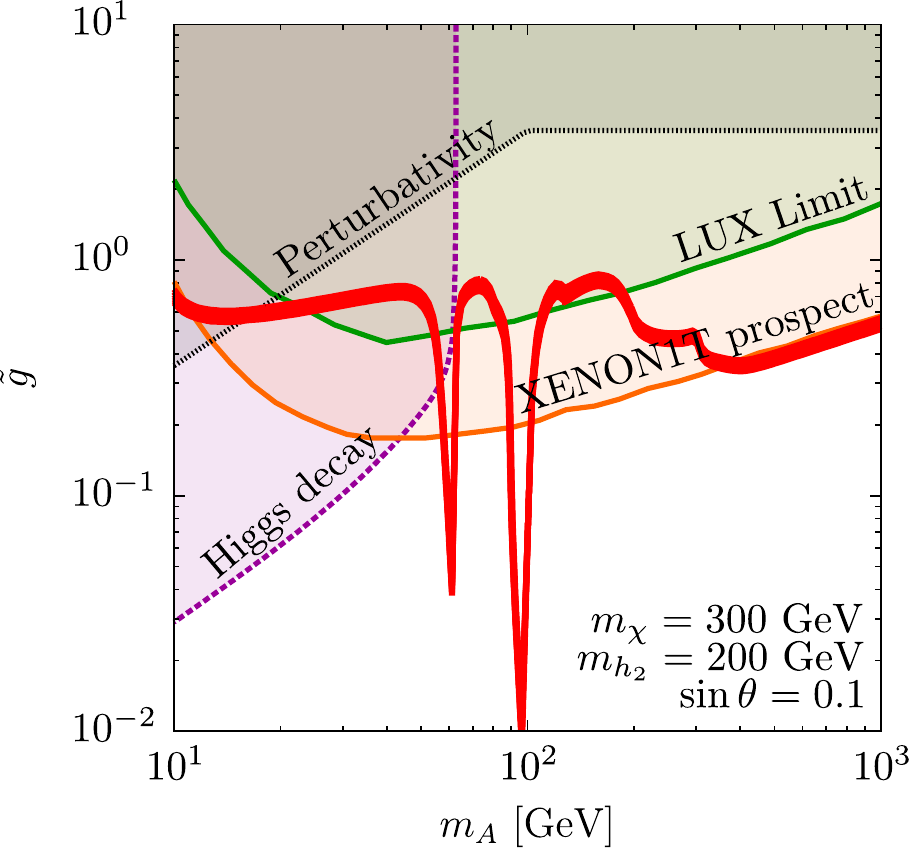}
\hspace{0.3cm}
\includegraphics[width=6.8 cm]{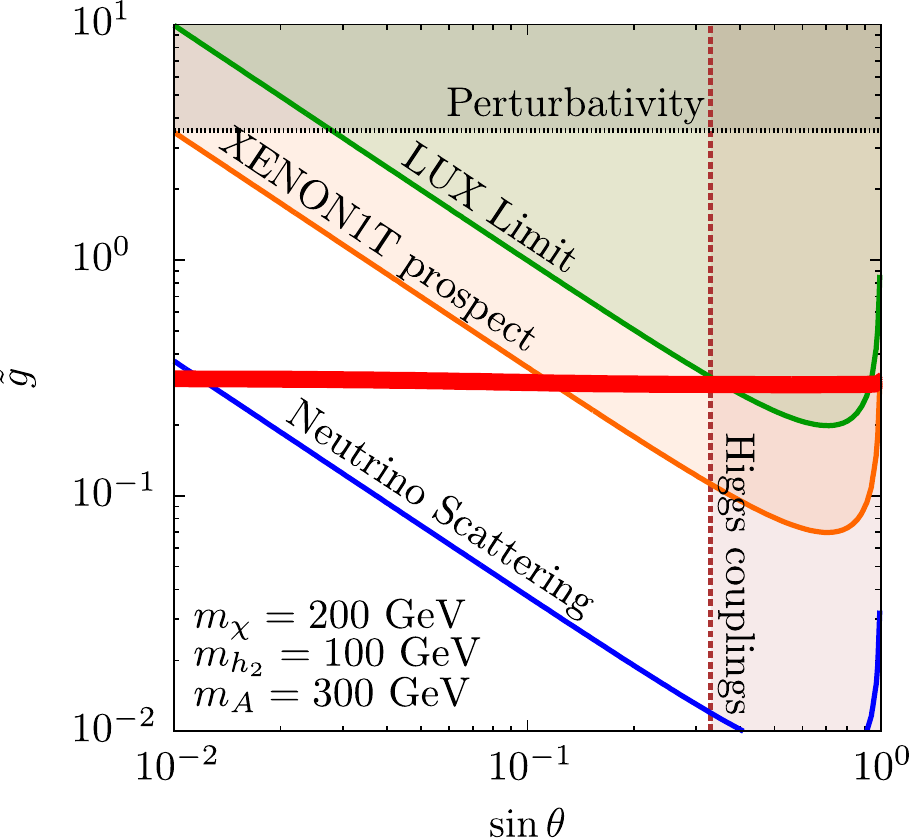}
\includegraphics[width=6.8 cm]{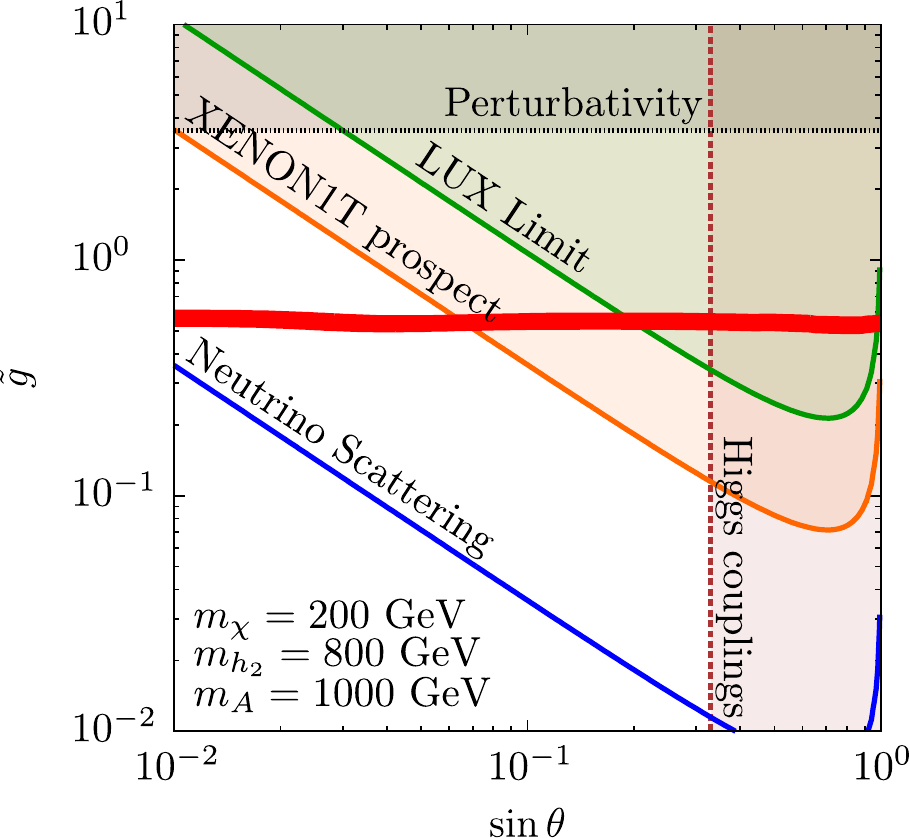}
\end{center}
\caption{ Dark matter constraints
 in the plane $(m_A,\tilde{g})$ (upper panels) and $(\sin\theta,\tilde{g})$ (lower panels) for SU(3) DM. 
The red band indicates the correct relic DM density. The other curves mark the following constraints:
grey -- perturbativity, purple -- invisible Higgs decay, dark red -- Higgs couplings, green -- LUX 2016 direct DM detection, orange -- XENON1T direct DM detection prospects. The blue line represents the direct detection event rate corresponding to the relic neutrino scattering off nuclei.
 }
\label{fig:dm2}
\end{figure}

 Our numerical results are shown in Fig.~\ref{fig:dm2}.
At low $m_A$, the dominant annihilation channel is $AA\rightarrow 
A^{3\;\prime}A^{3\;\prime}$ since $A^{3\;\prime}$ is always lighter than $A$, while for heavier DM, the $h_2 h_2$ and $\chi \chi$ 
final states become important. It is noteworthy that $m_A $ can be as low as 22~GeV at $\sin \theta=0.01$ without violating any constraints.
This is in contrast to the U(1) case where there is no analog of the
process $AA\rightarrow 
A^{3\;\prime}A^{3\;\prime}$, which excludes very light DM.
For $\sin\theta=0.1$, the effect of the channel $AA\rightarrow h_2 h_2$ or $AA\rightarrow \chi \chi$
is crucial to evade the LUX constraint so that the lowest allowed
$m_A$ is about 200 GeV (except for the resonance regions). The lower panels show that again the upper bound on $\sin\theta$ is of order $10^{-1}$ for electroweak masses.

This analysis can be repeated for larger SU(N) groups (see \cite{Gross:2015cwa}). In the minimal setting, the gauge symmetry is broken by VEVs of $N-1$ scalar $N$--plets. The lightest stable fields correspond to an SU(2) subgroup which gets broken at the last stage. The analog of $A^{3\;\prime}$ is lighter than the analogs of $A^{1,2}$ since it
mixes with the other Cartan generators. Thus many features of our analysis generalize to SU(N).

\section{Conclusion}

In the framework of Higgs portal dark matter, we have studied the interplay between the direct DM detection rates and the DM annihilation cross section. Focusing on spin-1 dark matter, we point out that the framework generally contains unstable states lighter than DM, which
have a significant coupling to the latter. This opens up an efficient DM annihilation channel into such states thereby effectively decoupling the annihilation cross section from the direct detection rate.
 The latter is suppressed as long as the mixing angle between the SM Higgs and the ``hidden Higgs'' is small. This allows us to circumvent the strong LUX/PANDA constraints while retaining the WIMP nature of
 our dark matter candidate.

 We have illustrated this mechanism with both Abelian and non--Abelian vector DM examples. In the U(1) and SU(2) cases, the ``hidden Higgs''
 must be lighter than DM to allow for efficient annihilation. For SU(3) and larger groups, the spectrum automatically contains light unstable
 vectors which provide DM with an annihilation channel. This argument assumes that CP symmetry is broken in the scalar sector, which in general is the case.
 
 The required mixing angle $\theta$ is of order $10^{-1}$ for ${\cal O}(100)$ GeV DM masses. A somewhat smaller $\theta$ (by a factor of a few) would suppress the direct detection rate beyond the reach of XENON1T. Dark matter can also be quite light, below 100 GeV, in which
 case the mixing angle is constrained to be   in the range $10^{-2}$ \ldots $10^{-1}$, depending on $m_A$.

We emphasize that the mechanism does not require any significant fine-tuning. When one of the hidden states turns lighter than DM,
the annihilation process becomes efficient immediately. The required hidden sector gauge coupling lies in the range $0.1 \ldots 1$ which
appears rather natural. 

\vspace{10pt}
\noindent
{\bf Acknowledgements} 

\noindent
C.G. and O.L. acknowledge support from the Academy of Finland, project {\it The Higgs Boson and the Cosmos}. 
S.P. is supported by the National Science Centre, Poland, under research grants
DEC-2014/15/B/ST2/02157, DEC-2015/19/B/ST2/02848
and DEC-2015/18/M/ST2/00054.
T.T. acknowledges support from JSPS Fellowships for Research Abroad.


\begin{thebibliography}{}

%\cite{Silveira:1985rk}
\bibitem{Silveira:1985rk} 
 V.~Silveira and A.~Zee,
 %``Scalar Phantoms,''
 Phys.\ Lett.\ {\bf 161B}, 136 (1985).
  %doi:10.1016/0370-2693(85)90624-0
  %%CITATION = doi:10.1016/0370-2693(85)90624-0;%%
  %389 citations counted in INSPIRE as of 31 Oct 2016


%\cite{Schabinger:2005ei}
\bibitem{Schabinger:2005ei} 
  R.~M.~Schabinger and J.~D.~Wells,
  %``A Minimal spontaneously broken hidden sector and its impact on Higgs boson physics at the large hadron collider,''
  Phys.\ Rev.\ D {\bf 72}, 093007 (2005)
  %doi:10.1103/PhysRevD.72.093007
  [hep-ph/0509209].
  %%CITATION = doi:10.1103/PhysRevD.72.093007;%%
  %193 citations counted in INSPIRE as of 16 Nov 2016

%\cite{Patt:2006fw}
\bibitem{Patt:2006fw}
 B.~Patt and F.~Wilczek,
 %``Higgs-field portal into hidden sectors,''
 hep-ph/0605188.
 %%CITATION = HEP-PH/0605188;%%



\bibitem{Hambye:2008bq} 
  T.~Hambye,
  %``Hidden vector dark matter,''
  JHEP {\bf 0901}, 028 (2009)
  %doi:10.1088/1126-6708/2009/01/028
  [arXiv:0811.0172 [hep-ph]].
  %%CITATION = doi:10.1088/1126-6708/2009/01/028;%%
  %122 citations counted in INSPIRE as of 16 Nov 2016

%\cite{Lebedev:2011iq}
\bibitem{Lebedev:2011iq} 
  O.~Lebedev, H.~M.~Lee and Y.~Mambrini,
  %``Vector Higgs-portal dark matter and the invisible Higgs,''
  Phys.\ Lett.\ B {\bf 707}, 570 (2012)
 % doi:10.1016/j.physletb.2012.01.029
  [arXiv:1111.4482 [hep-ph]].
  %%CITATION = doi:10.1016/j.physletb.2012.01.029;%%
  %95 citations counted in INSPIRE as of 16 Nov 2016


%\cite{Gross:2015cwa}
\bibitem{Gross:2015cwa} 
  C.~Gross, O.~Lebedev and Y.~Mambrini,
  %``Non-Abelian gauge fields as dark matter,''
  JHEP {\bf 1508}, 158 (2015)
  %doi:10.1007/JHEP08(2015)158
  [arXiv:1505.07480 [hep-ph]].
  %%CITATION = doi:10.1007/JHEP08(2015)158;%%
  %15 citations counted in INSPIRE as of 16 Nov 2016


%\cite{Arcadi:2016kmk}
\bibitem{Arcadi:2016kmk} 
  G.~Arcadi, C.~Gross, O.~Lebedev, Y.~Mambrini, S.~Pokorski and T.~Toma,
  %``Multicomponent Dark Matter from Gauge Symmetry,''
  arXiv:1611.00365 [hep-ph].
  %%CITATION = ARXIV:1611.00365;%%


%\cite{Akerib:2015rjg}
\bibitem{Akerib:2015rjg} 
  D.~S.~Akerib {\it et al.} [LUX Collaboration],
  %``Improved Limits on Scattering of Weakly Interacting Massive Particles from Reanalysis of 2013 LUX Data,''
  Phys.\ Rev.\ Lett.\  {\bf 116}, no. 16, 161301 (2016)
  %doi:10.1103/PhysRevLett.116.161301
  [arXiv:1512.03506 [astro-ph.CO]].
  %%CITATION = doi:10.1103/PhysRevLett.116.161301;%%
  %205 citations counted in INSPIRE as of 16 Nov 2016
  
  %\cite{Tan:2016zwf}
\bibitem{Tan:2016zwf} 
  A.~Tan {\it et al.} [PandaX-II Collaboration],
  %``Dark Matter Results from First 98.7 Days of Data from the PandaX-II Experiment,''
  Phys.\ Rev.\ Lett.\  {\bf 117}, no. 12, 121303 (2016)
 % doi:10.1103/PhysRevLett.117.121303
  [arXiv:1607.07400 [hep-ex]].
  %%CITATION = doi:10.1103/PhysRevLett.117.121303;%%
  %61 citations counted in INSPIRE as of 16 Nov 2016


%\cite{Pospelov:2007mp}
\bibitem{Pospelov:2007mp} 
  M.~Pospelov, A.~Ritz and M.~B.~Voloshin,
  %``Secluded WIMP Dark Matter,''
  Phys.\ Lett.\ B {\bf 662}, 53 (2008)
 % doi:10.1016/j.physletb.2008.02.052
  [arXiv:0711.4866 [hep-ph]].
  %%CITATION = doi:10.1016/j.physletb.2008.02.052;%%
  %406 citations counted in INSPIRE as of 16 Nov 2016

%\cite{Pospelov:2008jd}
\bibitem{Pospelov:2008jd} 
  M.~Pospelov and A.~Ritz,
  %``Astrophysical Signatures of Secluded Dark Matter,''
  Phys.\ Lett.\ B {\bf 671}, 391 (2009)
 % doi:10.1016/j.physletb.2008.12.012
  [arXiv:0810.1502 [hep-ph]].
  %%CITATION = doi:10.1016/j.physletb.2008.12.012;%%
  %356 citations counted in INSPIRE as of 16 Nov 2016

%\cite{LopezHonorez:2012kv}
\bibitem{LopezHonorez:2012kv} 
  L.~Lopez-Honorez, T.~Schwetz and J.~Zupan,
  %``Higgs portal, fermionic dark matter, and a Standard Model like Higgs at 125 GeV,''
  Phys.\ Lett.\ B {\bf 716}, 179 (2012)
  %doi:10.1016/j.physletb.2012.07.017
  [arXiv:1203.2064 [hep-ph]].
  %%CITATION = doi:10.1016/j.physletb.2012.07.017;%%
  %111 citations counted in INSPIRE as of 16 Nov 2016



%\cite{Farzan:2012hh}
\bibitem{Farzan:2012hh} 
  Y.~Farzan and A.~R.~Akbarieh,
  %``VDM: A model for Vector Dark Matter,''
  JCAP {\bf 1210}, 026 (2012)
 % doi:10.1088/1475-7516/2012/10/026
  [arXiv:1207.4272 [hep-ph]].
  %%CITATION = doi:10.1088/1475-7516/2012/10/026;%%
  %31 citations counted in INSPIRE as of 16 Nov 2016

%\cite{Baek:2012se}
\bibitem{Baek:2012se} 
  S.~Baek, P.~Ko, W.~I.~Park and E.~Senaha,
  %``Higgs Portal Vector Dark Matter : Revisited,''
  JHEP {\bf 1305}, 036 (2013)
 % doi:10.1007/JHEP05(2013)036
  [arXiv:1212.2131 [hep-ph]].
  %%CITATION = doi:10.1007/JHEP05(2013)036;%%
  %61 citations counted in INSPIRE as of 16 Nov 2016
  
  %\cite{Duch:2015jta}
\bibitem{Duch:2015jta} 
  M.~Duch, B.~Grzadkowski and M.~McGarrie,
  %``A stable Higgs portal with vector dark matter,''
  JHEP {\bf 1509}, 162 (2015)
  %doi:10.1007/JHEP09(2015)162
  [arXiv:1506.08805 [hep-ph]].
  %%CITATION = doi:10.1007/JHEP09(2015)162;%%
  %11 citations counted in INSPIRE as of 16 Nov 2016

%\cite{Kanemura:2010sh}
\bibitem{Kanemura:2010sh} 
  S.~Kanemura, S.~Matsumoto, T.~Nabeshima and N.~Okada,
  %``Can WIMP Dark Matter overcome the Nightmare Scenario?,''
  Phys.\ Rev.\ D {\bf 82}, 055026 (2010)
  %doi:10.1103/PhysRevD.82.055026
  [arXiv:1005.5651 [hep-ph]].
  %%CITATION = doi:10.1103/PhysRevD.82.055026;%%
  %133 citations counted in INSPIRE as of 16 Nov 2016


%\cite{Djouadi:2011aa}
\bibitem{Djouadi:2011aa} 
  A.~Djouadi, O.~Lebedev, Y.~Mambrini and J.~Quevillon,
  %``Implications of LHC searches for Higgs--portal dark matter,''
  Phys.\ Lett.\ B {\bf 709}, 65 (2012)
  %doi:10.1016/j.physletb.2012.01.062
  [arXiv:1112.3299 [hep-ph]].
  %%CITATION = doi:10.1016/j.physletb.2012.01.062;%%
  %262 citations counted in INSPIRE as of 16 Nov 2016
  
  %\cite{Gross:1984dd}
\bibitem{Gross:1984dd}
  D.~J.~Gross, J.~A.~Harvey, E.~J.~Martinec and R.~Rohm,
  %``The Heterotic String,''
  Phys.\ Rev.\ Lett.\  {\bf 54} (1985) 502.
%  doi:10.1103/PhysRevLett.54.502
  %%CITATION = doi:10.1103/PhysRevLett.54.502;%%
  

%\cite{Belanger:2014vza}
\bibitem{Belanger:2014vza} 
  G.~Belanger, F.~Boudjema, A.~Pukhov and A.~Semenov,
  %``micrOMEGAs4.1: two dark matter candidates,''
  Comput.\ Phys.\ Commun.\  {\bf 192}, 322 (2015)
  %doi:10.1016/j.cpc.2015.03.003
  [arXiv:1407.6129 [hep-ph]].
  %%CITATION = doi:10.1016/j.cpc.2015.03.003;%%
  %104 citations counted in INSPIRE as of 16 Nov 2016
  
  %\cite{Khachatryan:2016vau}
\bibitem{Khachatryan:2016vau}
  G.~Aad {\it et al.} [ATLAS and CMS Collaborations],
  %``Measurements of the Higgs boson production and decay rates and constraints on its couplings from a combined ATLAS and CMS analysis of the LHC pp collision data at $ \sqrt{s}=7 $ and 8 TeV,''
  JHEP {\bf 1608} (2016) 045
%  doi:10.1007/JHEP08(2016)045
  [arXiv:1606.02266 [hep-ex]].
  %%CITATION = doi:10.1007/JHEP08(2016)045;%%


%\cite{Khoze:2014xha}
\bibitem{Khoze:2014xha} 
  V.~V.~Khoze, C.~McCabe and G.~Ro,
  %``Higgs vacuum stability from the dark matter portal,''
  JHEP {\bf 1408}, 026 (2014)
  %doi:10.1007/JHEP08(2014)026
  [arXiv:1403.4953 [hep-ph]].
  %%CITATION = doi:10.1007/JHEP08(2014)026;%%
  %66 citations counted in INSPIRE as of 16 Nov 2016

%\cite{Karam:2015jta}
\bibitem{Karam:2015jta} 
  A.~Karam and K.~Tamvakis,
  %``Dark matter and neutrino masses from a scale-invariant multi-Higgs portal,''
  Phys.\ Rev.\ D {\bf 92}, no. 7, 075010 (2015)
  %doi:10.1103/PhysRevD.92.075010
  [arXiv:1508.03031 [hep-ph]].
  %%CITATION = doi:10.1103/PhysRevD.92.075010;%%
  %22 citations counted in INSPIRE as of 16 Nov 2016


%\cite{Karam:2016rsz}
\bibitem{Karam:2016rsz} 
  A.~Karam and K.~Tamvakis,
  %``Dark Matter from a Classically Scale-Invariant $SU(3)_X$,''
  Phys.\ Rev.\ D {\bf 94}, no. 5, 055004 (2016)
  %doi:10.1103/PhysRevD.94.055004
  [arXiv:1607.01001 [hep-ph]].
  %%CITATION = doi:10.1103/PhysRevD.94.055004;%%
  %5 citations counted in INSPIRE as of 16 Nov 2016


\end{thebibliography}
\end{document}